\documentclass{aa}
\usepackage[dvips,xdvi]{graphics}

\begin{document}
\newcommand{\MIC}{\mbox{$\mu$m}}
\newcommand{\MOL}{\mbox{$M/L{\rm _H}$}}
\newcommand{\HA}{\mbox{H$\alpha$}}
\newcommand{\EW}{\mbox{$W_\lambda$}}
\newcommand{\LSUN}{\mbox{$L_\odot$}}
\newcommand{\MSUN}{\mbox{$M_\odot$}}

\title{Starbursts in active galaxy nuclei: observational constraints from
IR stellar absorption lines
\thanks{Based on observations collected at the European Southern
Observatory, La Silla, Chile.}
}


\author
{
E. Oliva\inst{1}
\and
L. Origlia\inst{2}
\and 
R. Maiolino\inst{1}
\and
A.F.M. Moorwood\inst{3}
}

\institute{ 
Osservatorio di Arcetri, Largo E. Fermi 5, I-50125 Firenze, Italy
\and
Osservatorio Astronomico di Bologna, Via Ranzani 1, I--40127 Bologna, Italy
\and
European Southern Observatory, Karl Schwarzschild Str. 2, D-85478 Garching, 
Germany
}

\offprints{E. Oliva}
             
\thesaurus{  03(
              11.01.2;          
               11.19.1;          
               11.19.3;          
               11.19.5           
          }

\date{ Received 13 April 1999/ Accepted ...}
\maketitle

\begin{abstract} 
High quality infrared spectra of active galaxies including the
stellar absorption features of Si at 1.59 \MIC, CO(6,3) at 
1.62 \MIC, and CO(2,0) 
at 2.29 \MIC\  are used to measure  the stellar
mass to light ratio at 1.65 \MIC\   (\MOL) and
investigate the occurrence of circum--nuclear starbursts.
We find that old and powerful starbursts are relatively common
in obscured AGNs (5 objects out of 13) while absent in genuine Seyfert 1's 
(0 objects out of 8). 

The data are also used to derive the non--stellar continuum which is very red
and compatible with emission from warm ($\simeq$1000 K) dust even in bare
Sy1's, thus indicating that the near IR nuclear continuum is reprocessed
radiation.
Hot dust ($\simeq$800 K) emission is also detected in a few 
obscured AGNs, including the Seyfert 2 prototype NGC1068.
The observed non--stellar flux is too high to be accounted for by
scattered light and therefore indicates that the material obscuring the AGN
must have a quite small ($\la$1 pc)
projected size.

\end{abstract}

\section{Introduction} 

Starburst events may recurrently occur in the central region of galaxies,
over-imposed to an old, quiescent stellar population such as that 
found in normal galaxies.
Estimating the starburst parameters -- i.e. their actual age, duration,
temporal and spatial
evolution, the shape of the initial mass function etc. -- is a very 
difficult and controversial task even in 
the ``cleanest'' and best studied objects.
In Seyfert galaxies the situation is further complicated by the presence of
the active nucleus which could out-shine and hide the emission from
the starburst at most wavelengths. Although 
some evidence of circum--nuclear starbursts has been occasionally
reported in a few type 2 Seyferts, it is not yet clear if such events
are common or just incidentally  related to the AGN
 (see e.g. the recent review by Moorwood 
\cite{moorwood96} and references therein).
 
Absorption features in near infrared spectra of galaxies  
can be used to study the cool stellar population whose
physical properties mainly depend on age and
metallicity. In a starburst, red supergiant and intermediate mass 
AGB stars dominate
the IR stellar light from $\sim\!10^7$ up to a few $\times10^9$ yr
while in old stellar systems the emission is dominated by low mass
red giants evolving on the RGB.
In a series of previous papers (Origlia et al. 1993, hereafter \cite{OMO93},
Origlia et al. \cite{origlia97}, Oliva \& Origlia \cite{origlia98}) 
we investigated the behaviour
of a carefully selected set of stellar absorption features -- namely 
Si 1.59 \MIC, CO(6,3) 1.62 \MIC\  and CO(2,0) 2.29 \MIC\ -- 
with stellar parameters
such as temperature, gravity, microturbulence velocity and chemical
composition. We found, in particular, that the equivalent width of the
CO(6,3) second overtone provides a powerful
tool to measure the metallicity of stellar systems. 
\begin{figure}
\centerline{\resizebox{8.8cm}{!}{\rotatebox{0}{\includegraphics{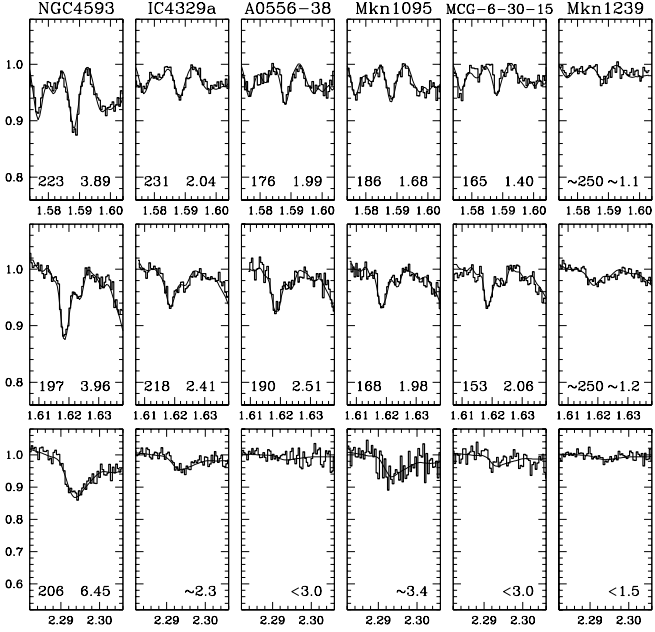}}}}
\caption{
Normalized spectra of type 1 Seyferts centered at the positions 
of the stellar
absorption features of Si 1.59 \MIC\  (top panels), CO(6,3) 1.62 \MIC\
(central panels) and CO(2,0) 2.29 \MIC\  (bottom panels).
Wavelengths are in the rest frame and the thin lines show the 
results of the spectral fit used to determine 
equivalent widths  and velocity dispersions  which are listed
in Table~\ref{tabew}.
}
\label{figsy1}
\end{figure}

\begin{figure}[!]
\centerline{\resizebox{8.8cm}{!}{\rotatebox{0}{\includegraphics{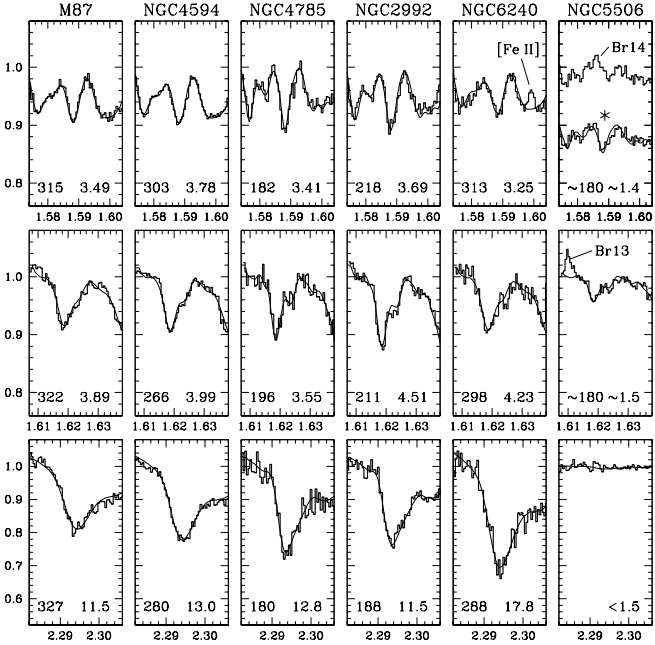}}}}
\caption{
Same as Fig.~\ref{figsy1} for obscured AGNs.
The top right panel also includes the spectrum of NGC5506 corrected for
the Br14 hydrogen emission line which fills in the Si 1.59 feature.
See the $2^{nd}$ paragraph
of Sect.~\ref{results} where we also discuss the nebular
line of [FeII] at 1.5995 \MIC\ which is visible in the spectrum of NGC6240.
}
\label{figsy2}
\end{figure}

In Oliva et al. (1995, hereafter \cite{paper1}) we applied our
IR diagnostics to a limited sample of normal and active galaxies and
used the velocity dispersion of the stellar absorption features to estimate
the dynamical mass and hence the
mass to light ratio of the dominant stellar population.
This provided evidence for starburst events  in several type 2, 
but not in type 1 Seyferts.
This paper is a follow--up of \cite{paper1} including
an enlarged sample of Seyferts. 
The data are presented in Sect.~\ref{observations} and the results
are discussed in Sect.~\ref{results}. In Sect.~\ref{conclusion} 
we draw our conclusions.

%
%

\section{Observations and data reduction}
\label{observations}

The spectra were collected in March 1995 at the ESO-NTT telescope using the
IRSPEC infrared spectrometer (Moorwood et al. \cite{moorwood91}, 
Gredel \& Weilenmann \cite{gredel92}). 
Instrument setup and observational procedures were as in \cite{paper1}.
Spectra were taken using a slit width of 4.4$\arcsec$ or 2 pixels  which
yields a resolving power of
$R\simeq$1600 and 2500 at 1.6 and 2.3 \MIC, respectively.
Each object was observed with the grating set at 
1.59, 1.62 and 2.29 $\mu$m (rest wavelengths)
to include the stellar absorption  features of interest.
Each 2D spectrum consisted of several ABBA cycles
(A--object, B--sky or, for compact sources, object moved
$\simeq$30$\arcsec$ along the slit) each exposure being the mean of
two 60 second frames. Total integration time per object varied between
1 and 4 hours, depending on the brightness of the source.
Data were spectroscopically calibrated (i.e. corrected
for instrumental and  atmospheric transmission) using spectra of O5/O6
main sequence stars taken at the same grating positions as the objects.
Absolute fluxes were derived from observations of photometric standard stars.
The 1D spectra were extracted from the central 3 rows of the array.
Data reduction was performed using the IRSPEC context in  MIDAS
the standard ESO reduction package.
More details can be found in Oliva \& Origlia
\cite{OHsky}, \cite{paper1} and in the MIDAS manual.

Velocity broadening ($\sigma$=FWHM/2.35) and equivalent width
of the stellar absorption features were determined using
the spectral fitting procedure described in \cite{paper1}.
This best--fits the data using broadened stellar spectra taken 
with the same instrument and also includes a free parameter representing
a featureless non--stellar continuum diluting the features.
More details can be found in Sect. 2.1 of \cite{paper1}.

\section{ Results and discussion }
\label{results}
\begin{table*}
\caption{ Observed parameters}
\label{tabew}
{
\def\UNO{\mbox{$^{(1)}$}}
\def\DUE{\mbox{$^{(2)}$}}
\def\TRE{\mbox{$^{(3)}$}}
\def\A{\mbox{$^a$}}
\def\B{\mbox{$^b$}}
\def\C{\mbox{$^c$}}
\def\D{\mbox{$^d$}}
\def\E{\mbox{$^e$}}
\def\F{\mbox{$^f$}}
\def\SIM{\llap{$\sim$}}
\def\UL{\llap{$<$}}
\def\tablerule{\noalign{\hrule}}
\def\SKIP{&& && && && && && && && && &\cr\noalign{\vglue-6pt}}
\def\BOT{\noalign{\vglue-6pt}&& && && && && && && && && &\cr\tablerule}
\hbox{\vbox{\tabskip=0pt\offinterlineskip
\halign to\hsize{
\strut#& \vrule #\tabskip=.2em plus3em&
#\hfil&#\vrule&       
\hfil#\hfil&#&        
\hfil#\hfil&#&        
\hfil#\hfil&#\vrule&  
\hfil#\hfil&#&        
\hfil#\hfil&#&        
\hfil#\hfil&#\vrule&  
\hfil#\hfil&#&        
\hfil#\hfil&\vrule    
# \tabskip=0pt\cr
\tablerule\SKIP
&& && \multispan5 \hfil Velocity dispersions\UNO \hfil &&
   \multispan5 \hfil Equivalent widths\DUE \hfil &&
   \multispan3 \hfil Fluxes\TRE \hfil &\cr
&& Object  && $\sigma(1.59)$  && $\sigma(1.62)$ && $\sigma(2.29)$ 
           && $W_\lambda$(1.59) && $W_\lambda$(1.62) && $W_\lambda$(2.29)
           && $F_{-11}$(1.62) && $F_{-11}$(2.29) &\cr
\SKIP\tablerule\SKIP
&& {\it Seyfert 1's} && && && && && && && && &\cr
&& NGC~4593 && 223 && 197 && 206 && 3.9 && 4.0 && 6.5 && 5.7 && 3.3 &\cr
&& IC~4329a && 231 && 218 && --- && 2.0 && 2.4 && \SIM2 && 7.6 && 6.6 &\cr
&& Mkn~1095 && 186 && 168 && --- && 1.7 && 2.0 && \SIM3 && 2.4 && 1.8 &\cr
&& Mkn~1239\A  && \SIM250 && \SIM250 && --- 
                   && \SIM1.1 && \SIM1.2 && \UL1 && 5.3 && 5.5 &\cr
&& A~0556-38 && 176 && 190 && --- && 2.0 && 2.5 && \UL3 && 2.6 && 2.5 &\cr
&& MCG-6-30-15 && 165 && 153 && --- && 1.4 && 2.1 && \UL3 && 4.4 && 3.5 &\cr
\SKIP
&& {\it Obscured AGNs$^{\;b}$} && && && && && && && && &\cr
&& NGC~2992 && 218 && 211 && 188 && 3.7 && 4.5 && 11.5 && 5.8 && 2.7 &\cr
&& NGC~4594\C && 303 && 266 && 281 && 3.8 && 4.0 && 13.0 && 32. && 11. &\cr
&& NGC~4785 && 182 && 196 && 180 && 3.4 && 3.6 && 12.8 && 2.8 && 1.1 &\cr
&& NGC~5506\D && \SIM180 && \SIM180 && --- && \SIM1.4 && \SIM1.5 && \UL1.5
                                                    && 6.5 && 8.8 &\cr
&& NGC~6240\E && 313 && 298 && 288 && 3.3 && 4.2 && 17.8 && 4.0 && 2.3 &\cr
&& M~87\F   && 315 && 322 && 327 && 3.5 && 3.9 && 11.5 && 11. && 3.8 &\cr
\BOT }}}

\smallskip
\def\NOTA#1#2
{\hbox{\vtop{\hbox{\hsize=0.015\hsize\vtop{#1}}}
      \vtop{\hbox{\hsize=0.97\hsize\vtop{#2}}}}}
 
\NOTA{\UNO}{ Velocity dispersions (km/s) of the stellar absorption lines
Si 1.59 $\mu$m, CO(6,3) 1.62 $\mu$m, CO(2,0) 2.29 $\mu$m.
Typical errors are $\pm$20 km/s.}
 
\NOTA{\DUE}{ Equivalent widths (\AA) of above stellar lines.
Typical errors are 0.3 \AA\ for \EW(1.59), 
\EW(1.62) and 0.6 \AA\  for \EW(2.29).}
 
\NOTA{\TRE}{ Observed continuum flux, units of
10$^{-11}$ erg cm$^{-2}$ s$^{-1}$ $\mu$m$^{-1}$.}
 
\vskip2pt
\NOTA{$\;^a$}{Continuum of Mkn~1239 is dominated by non--stellar emission,
stellar features are shallow (see Fig.~\ref{figsy1}).}

\NOTA{$\;^b$}{Objects with an obscured active nucleus,
classified from optical and/or X--ray spectra (see also
Sect.~\ref{individual})}
 
\NOTA{$\;^c$}{Optical stellar lines yield $\sigma$=256 km/s
(Whitmore et al. 1985).}
 
\NOTA{$\;^d$}{Non-stellar continuum and hydrogen emission lines are both
strong in NGC~5506 (see Fig.~\ref{figsy2} and Sects.~\ref{results},
 \ref{ngc5506}). }
 
\NOTA{$\;^e$}{ 
Infrared spectra of CO(2,0) by Lester \& Gaffney (\cite{lester94}),
Doyon et al. (\cite{doyon94}) and Shier et al. (\cite{shier96}) yield
$\sigma$(2.29)=350 km/s. This galaxy is also discussed
in Sect.~\ref{ngc6240}}
 
\NOTA{$\;^f$}{Radio galaxy, optical stellar lines yield $\sigma$=335 km/s
(Whitmore et al. 1985).}
 
}
\end{table*}

The normalized spectra are displayed in Figs.~\ref{figsy1}, \ref{figsy2}
 together with the
fitted profiles. Continuum fluxes are listed in Table~\ref{tabew}
together with the velocity dispersions and equivalent widths of the
stellar features derived from the spectral fitting. 
As in \cite{paper1},
the scatter between the velocity dispersions of the various
features in each object indicates that the 
errors should not exceed $\pm$20 km/s (equivalent to about half a pixel).
Typical errors on equivalent widths are $\pm$0.3 \AA\  for Si and CO(6,3)
and $\pm$0.6 \AA\ for CO(2,0), these are mainly determined by the accuracy of
continuum positioning and by the correction for velocity broadening
described in Sect.~2.1 of \cite{paper1}. \\

Spectral fitting was particularly difficult in the case of NGC5506, a
Sy2 galaxy with unusually strong hydrogen emission lines such as
Br13 (1.6109 \MIC),
which is visible on the blue side of CO(6,3), and
Br14 (1.5880 \MIC), which fills in the Si 1.59 feature 
(see the top right panel of Fig.~\ref{figsy2}).
To recover the stellar absorption feature at 1.59 \MIC\  we took the
observed emission profile of Br13 and added it in absorption at the position 
of Br14 after correction for the intrinsic Br13/Br14=1.25 ratio.
The resulting spectrum is also displayed in Fig.~\ref{figsy2} (marked by an 
asterisk) and its shape is compatible with $\sigma\!\simeq\!180$ km/s, i.e.
the velocity dispersion derived from CO(6,3).
The only other object displaying nebular lines is NGC6240, an
interacting system with extremely strong [FeII] 1.644 and H$_2$ 2.121
\MIC\  emission lines
(e.g. van der Werf et al. \cite{paul93}). 
The satellite line of [FeII] at 1.5995 \MIC\  is clearly
detected in our spectra (see Fig.~\ref{figsy2}) and 
was therefore excluded by the spectral fitting.
The line
intensity is $\simeq\!5\times 10^{-15}$ erg cm$^{-2}$ s$^{-1}$ or about 3\%
of [FeII] 1.644 \MIC. This ratio indicates an electron density 
$\la\!10^3$ cm$^{-3}$ (see Fig. 7 of Oliva et al. \cite{oliva_r103}).

\begin{table*}
\caption{ Stellar fluxes, reddening and non--stellar continua.}
\label{tabcont}
{
\def\UNO{\mbox{$^{(1)}$}}
\def\DUE{\mbox{$^{(2)}$}}
\def\TRE{\mbox{$^{(3)}$}}
\def\QUA{\mbox{$^{(4)}$}}
\def\A{\rlap{$^a$}}
\def\B{\rlap{$^b$}}
\def\C{\rlap{$^c$}}
\def\D{\rlap{$^d$}}
\def\E{\rlap{$^e$}}
\def\SIM{\llap{$\sim$}}
\def\UL{\llap{$<$}}
\def\LL{\llap{$>$}}
\def\tablerule{\noalign{\hrule}}
\def\Skip{&& && && && && && && && && &\cr\noalign{\vglue-8pt}}
\def\SKIP{&& && && && && && && && && &\cr\noalign{\vglue-6pt}}
\def\BOT{\noalign{\vglue-6pt}&& && && && && && && && && &\cr\tablerule}
\hbox{\vbox{\tabskip=0pt\offinterlineskip
\halign to\hsize{
\strut#& \vrule #\tabskip=.2em plus3em&
#\hfil&#\vrule&       
#&#&        
#&#&        
\hfil#\hfil&#\vrule&  
\hfil#\hfil&#&        
\hfil#\hfil&#&        
\hfil#\hfil&#&        
\hfil#\hfil&#&        
\hfil#\hfil&\vrule    
# \tabskip=0pt\cr
\tablerule\SKIP
&& && \multispan5 \hfil Stellar\UNO \hfil &&
   \multispan9 \hfil Non--stellar\DUE \hfil &\cr
&& Object  && \hfil $F_{*}$(1.62) \hfil  && \hfil $F_{*}$(2.29)\hfil
            && ${\rm A_H}$\TRE
           && $F_{nuc}$(1.62) && $F_{nuc}$(2.29) 
	   && $\alpha$ && $T_{dust}$ && $L_{2.29}/L_{FIR}$\QUA &\cr

\SKIP\tablerule\SKIP
&& {\it Seyfert 1's}  && && && && && && && && &\cr
&& NGC~4593 && 5.7 \hfil (100\%) && 2.0 \hfil (60\%) && 0.0 
           && \UL0.6 && 1.3 && \LL4\ (2.8)\A && \UL900 && 0.17 &\cr
&& IC~4329a && 4.6 \hfil (60\%)  && 1.6 \hfil (25\%) && 0.0 
           && 3.0    && 5.0 && 3.5\ (0.7)\A && 940 &&  1.3 &\cr
&& Mkn~1095 && 1.2 \hfil (50\%) && .45 \hfil (25\%) &&  0.0
            && 1.2 && 1.4 && 2.4 && 1080  && 0.93 &\cr
&& Mkn~1239 && 1.6 \hfil (30\%) && \UL0.6 \hfil ($<$10\%) && 0.0
           && 3.7 && 4.9 && 2.8 && 1030 && 1.5 &\cr
&& MCG-6-30-15 && 1.8 \hfil (40\%) && \UL0.7 \hfil ($<$20\%) && \UL0.3 
           && 2.6 && 2.9 && 2.3\ (2.4)\A && 1100 && 1.3 &\cr
&& A~0556-38 && 1.4 \hfil (55\%) && \UL0.5 \hfil ($<$20\%) && 0.0 
           && 1.2 && 2.0 && 3.5 && 940  && 2.3 &\cr
\Skip
&& NGC~1365\B && 7.8 \hfil (65\%) && 2.9 \hfil (40\%) && 0.0
           && 4.2 && 4.4 && 2.1 && 1130  && 0.01 &\cr
&& NGC~3783\B && 3.3 \hfil (55\%) && \UL1.2 \hfil ($<$20\%) && 0.0
           && 2.7 && 4.6 && 3.6\ (1.8)\A && 930  && 0.63 &\cr
&& Fairall~9\B && 1.1 \hfil (50\%) && \UL0.4 \hfil ($<$25\%) &&  0.0
           && 1.1 && 1.2 && 2.4\ (2.3)\A && 1090  && -- &\cr
\SKIP
&& {\it Obscured AGNs$^{\;c}$}  && && && && && && && && &\cr
&& NGC~2992 && 5.8 \hfil (100\%) && 2.4 \hfil (90\%) && 0.4
    && \UL0.6 && \SIM0.3 && \LL0.0 &&  && \SIM0.01 &\cr  
  && && && && 
  && \SIM1.5\D\ && \SIM3\D\ && \SIM4\ (3.4)\A && \SIM900 
  && \SIM 0.1 &\cr 
&& NGC~4594 && 32. \hfil (100\%) && 11. \hfil (100\%) && 0.0
           && \UL3 && \UL1 && -- && -- && \UL0.08 &\cr
&& NGC~4785 && 2.8 \hfil (100\%) && 1.1 \hfil (100\%) && 0.3
           && \UL0.3 && \UL0.1 && -- && -- && \UL0.009 &\cr
&& NGC~5506 && 2.6 \hfil (40\%) && \UL1.3 \hfil ($<$15\%) && 0.6\E\ 
           && 3.9 && 7.7 && 4.0\ (3.5)\A && 890 && 0.46 &\cr
&& NGC~6240 && 4.0 \hfil (100\%) && 2.3 \hfil (100\%) &&  1.1
           && \UL0.4 && \UL0.5 && -- && --  && \UL0.01 &\cr
&& M~87     && 11. \hfil (100\%) && 3.8 \hfil (100\%) &&  0.0
           && \UL1 && \UL0.8 && -- && --  && \UL0.70 &\cr
\Skip
&& NGC~1052\B && 11. \hfil (100\%) && 4.0 \hfil (100\%) && 0.0
              && \UL1   && \UL0.8  && -- && --  &&  \UL0.38 &\cr
&& NGC~1068\B && 17. \hfil (65\%) && \UL6.6 \hfil ($<$20\%) &&  \UL0.3
           && 9.0 && 26. && 5.1 && 780  && 0.07 &\cr
&& NGC~2110\B && 7.3 \hfil (100\%) && 3.3 \hfil (100\%) &&  0.6
           && \UL0.8 && \UL0.3 && -- && --  && \UL0.03 &\cr
&& NGC~4945\B   && 6.6 \hfil (100\%) && 6.3 \hfil (100\%) && 2.3
              && \UL0.7 && \UL1.3  && -- && --  && \UL0.001&\cr
&& NGC~7582\B && 9.5 \hfil (100\%) && 4.9 \hfil (70\%) &&  0.9
           && \UL1 && 2.0 && \LL4\ (2.9)\A && \UL900 
           &&  0.02 &\cr
&& Circinus\B && 36. \hfil (100\%) && 18. \hfil (100\%) &&  0.8
           && \UL4 && \UL4 && -- && -- &&  \UL0.006 &\cr
&& A~0945-30\B && 7.0 \hfil (100\%) && 2.5 \hfil (60\%) &&  0.0
           && \UL0.8 && 1.6 && \LL4 && \UL900  && -- &\cr
\SKIP
&& {\it Starbursters}  && && && && && && && && &\cr
&& NGC~253\B    && 18. \hfil (100\%) && 11. \hfil (100\%) && 1.3
              && \UL2 && \UL2   && -- && --  && \UL0.001 &\cr
&& NGC~1614\B   && 4.7 \hfil (100\%) && 2.3 \hfil (100\%) && 0.8
              && \UL0.5 && \UL0.5  && -- && --  && \UL0.008 &\cr
&& NGC~1808\B   && 17. \hfil (100\%) && 7.3 \hfil (100\%) && 0.5
              && \UL2 && \UL1.5  && -- && -- &&  \UL0.008 &\cr
&& NGC~3256\B   && 5.2 \hfil (100\%) && 2.5 \hfil (100\%) && 0.7
              && \UL0.5 && \UL0.5  && -- && --  && \UL0.003 &\cr
&& NGC~7552\B   && 8.9 \hfil (100\%) && 4.0 \hfil (100\%) && 0.6
              && \UL0.9 && \UL0.8  && -- && --  && \UL0.005 &\cr
&& NGC~7714\B   && 3.0 \hfil (100\%) && 1.2 \hfil (100\%) && 0.3
              && \UL0.3 && \UL0.3  && -- && --  && \UL0.01 &\cr
\BOT }}}

\smallskip
\def\NOTA#1#2
{\hbox{\vtop{\hbox{\hsize=0.015\hsize\vtop{#1}}}
      \vtop{\hbox{\hsize=0.97\hsize\vtop{#2}}}}}

\NOTA{\UNO}{ Stellar flux in absolute units 
($\times$10$^{-11}$ erg cm$^{-2}$ s$^{-1}$ $\mu$m$^{-1}$)
and in percentage (in brackets) of the total observed flux, note
that ``100\%'' means $\ge$90\% at 1.62 $\mu$m and $\ge$80\% at 2.29 $\mu$m.
These were derived using the equivalent width diagrams (Figs.~\ref{figew1},
\ref{figew2}) discussed
in Sect.~\ref{nonstellar}. }

\NOTA{\DUE}{ Non--stellar flux in absolute units 
($\times$10$^{-11}$ erg cm$^{-2}$ s$^{-1}$ $\mu$m$^{-1}$)
and with shape parameterized in terms of a power law slope 
($F_\nu\propto\nu^{-\alpha}$) and 
temperature of gray dust emission ($T_{dust}$,
with $\kappa_\lambda\propto\lambda^{-1.5}$).}

\NOTA{\TRE}{ Extinction at H (1.65 $\mu$m) assuming an 
intrinsic stellar colour H--K=0.2 [i.e. $F$(1.62)/$F$(2.29)=2.9] 
and a reddening curve A$_\lambda\!\propto\!\lambda^{-1.8}$,
errors are $\approx$0.1 mag. Note that ``0.0'' effectively means $<$0.2.}

\NOTA{\QUA}{ $L_{2.29}=\nu L_\nu(2.29)$ is the luminosity of the
non--stellar component, $L(FIR)$ is the luminosity in the IRAS bands.
}

\vskip2pt
\NOTA{$\;^a$}{ Values in brackets are the power law slopes derived
by Kotilainen et al. (\cite{koti1}) using imaging decomposition
techniques. Note that their very flat indices for NGC3783 and IC4329a
are not confirmed by our data.}

\NOTA{$\;^b$}{ Data from \cite{paper1}}

\NOTA{$\;^c$}{ Objects with an obscured active nucleus,
classified from optical and/or X--ray spectra (see also
Sect.~\ref{individual})}

\NOTA{$\;^d$}{ The nucleus of NGC~2992 was observed in a low state. 
The higher fluxes are from the Feb 1988 observations of Kotilainen
et al. (1992) after subtraction of the stellar fluxes derived here.}

\NOTA{$\;^e$}{ Reddening in NGC~5506 from hydrogen recombination lines. }

}
\end{table*}

\begin{figure}
\centerline{\resizebox{8.8cm}{!}{\rotatebox{0}{\includegraphics{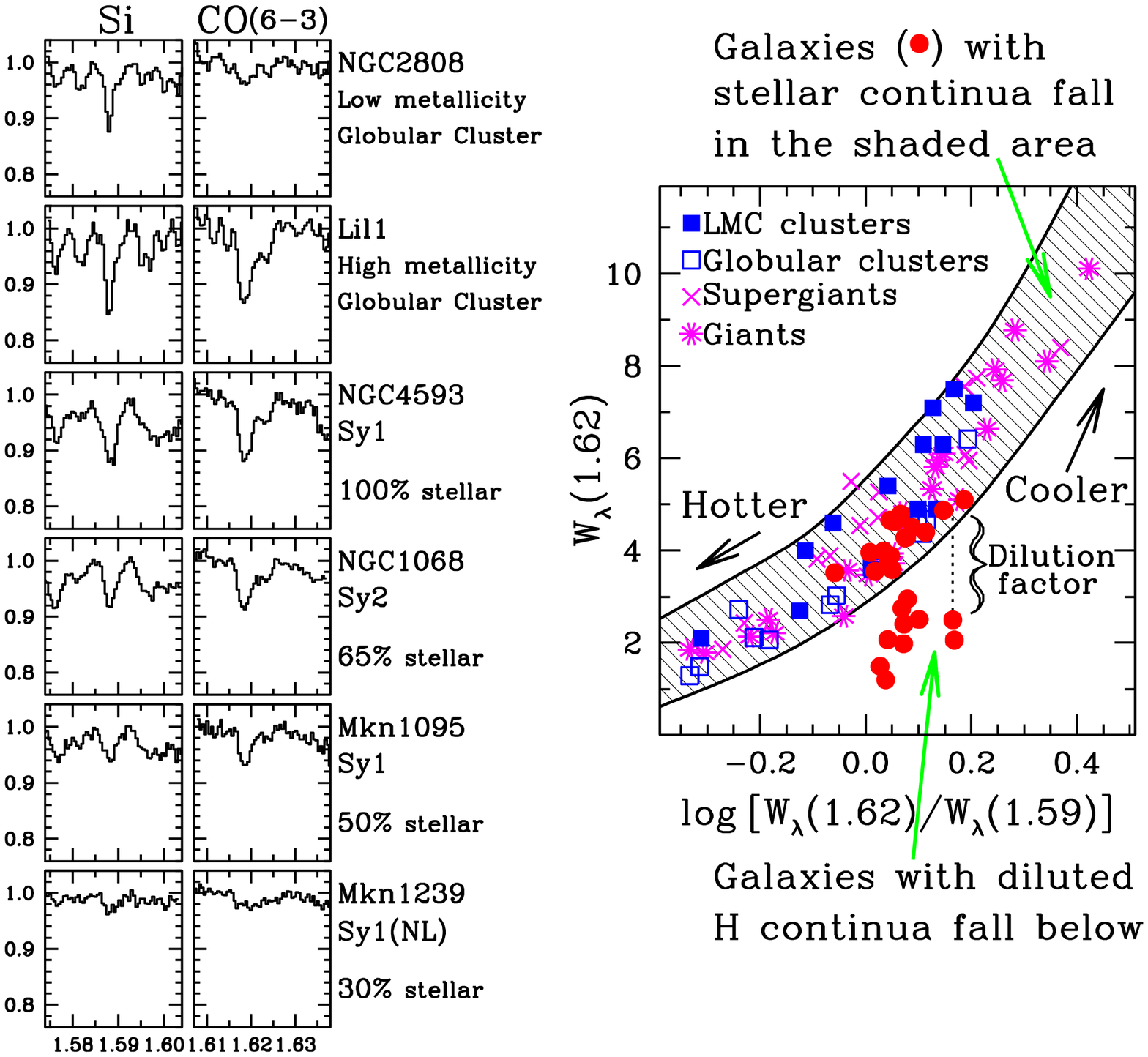}}}}
\caption{
The right hand panel shows the
\EW(1.62) vs. \EW(1.62)/\EW(1.59) 
diagram used to estimate the non--stellar continuum emission at 1.62
\MIC. Measurements of objects dominated by stellar light are marked
and normalized spectra of representative objects are shown in the left hand
panels. Galaxies with diluted stellar continua fall below the dashed line,
see Sect.~\ref{nonstellar} for details.
}
\label{figew1}
\end{figure}

\begin{figure}
\centerline{\resizebox{8.8cm}{!}{\rotatebox{0}{\includegraphics{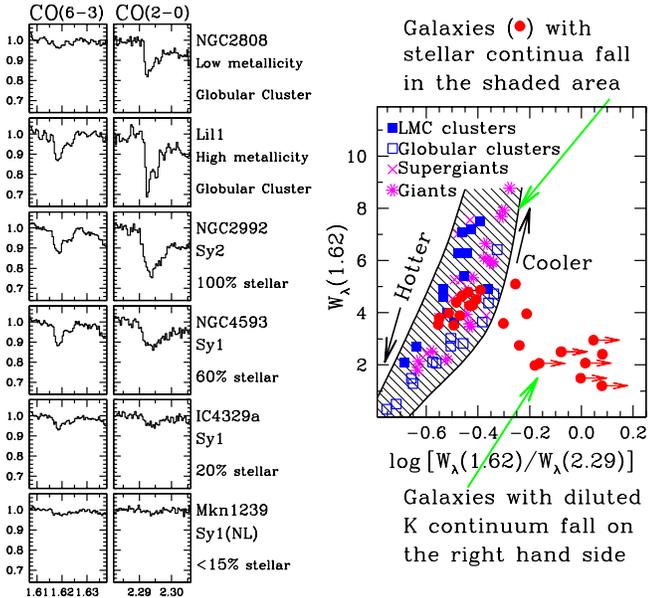}}}}
\caption{
Right hand panel:
\EW(1.62) vs. \EW(1.62)/\EW(2.29) 
diagram used to estimate the non--stellar continuum emission at 2.29
\MIC. Measurements of objects dominated by stellar light are marked
and normalized spectra of representative objects are shown in the left hand
panels. Galaxies with diluted stellar continua fall to the right hand side
of the dashed region,
see Sect.~\ref{nonstellar} for details.
}
\label{figew2}
\end{figure}

\subsection{ Non--stellar continuum emission}
\label{nonstellar}

The contribution of non--stellar emission to the observed continuum
flux at 1.6 \MIC\  can be estimated using the 
\EW(1.62) vs. \EW(1.62)/\EW(1.59)
plot of Fig.~\ref{figew1}. \\
The method relies on observations of stars and stellar clusters showing
that there is
a limited range of \EW(1.62) for a given \EW(1.62)/\EW(1.59) ratio, and
both quantities increase going to cool stars (\cite{OMO93}). Warm stars 
(e.g. low metallicity globular clusters) have shallow CO(6,3) and relatively
strong Si 1.59 and therefore lie at the bottom left of the diagram
of Fig.~\ref{figew1}. Going to lower temperatures (e.g. higher metallicity 
globular clusters) the strength of CO(6,3) increases much
more rapidly than Si 1.59
and the point moves toward the upper right of the diagram.
Objects with diluted stellar features lie below the locus occupied by stars
and star clusters because they have a shallower CO(6,3) index, while
\EW(1.62)/\EW(1.59) is not significantly affected by dilution since
the two features lie close in wavelength. 
The fraction of non--stellar continuum is simply given by the vertical
displacement of the point in the diagram.

The dilution at 2.29 $\mu$m is determined in a similar way from the
diagram in Fig.~\ref{figew2} where objects with significant non--stellar 
continua lie on the right hand side. 
The only complication is that one
first needs to correct \EW(1.62) for dilution before
determining the non--stellar fraction from the horizontal
displacement in the diagram (see also Fig. 2b of \cite{paper1}).\\

The effect of non--stellar emission is much more evident at the longer 
wavelengths where the CO(2,0) feature is weak or virtually absent in all
Sy1 spectra (see Fig.~\ref{figsy1} and Fig.~1c of \cite{paper1}).
Diluted features are less common in type 2 Seyferts, remarkable
exceptions being  
NGC1068 (see \cite{OMO93}) and NGC5506 (see Fig.~\ref{figsy2})
whose 2.29 \MIC\  spectra are basically featureless and therefore
dominated by non--stellar radiation.

The stellar features around 1.6 $\mu$m 
are much less diluted and clearly visible in all but one object (Mkn1239).
This implies that the diluting continuum is very red and its colour
temperature (second last column of Table~\ref{tabcont})
is compatible with emission by relatively hot dust ($\simeq$1000 K) even
in bare Sy1's such as NGC3783 and IC4329a.
We do not therefore confirm the results of Kotilainen et al. (\cite{koti1})
who, based on decomposition modelling of near IR images,
claim very flat nuclear continua in many Seyferts including several of those
in our sample
(see 7$^{th}$ column of Table~\ref{tabcont}).

The above results indicate that the nuclear continuum
at 1.6--2.3 \MIC\  is due to reprocessed radiation, i.e.
emission from dusty clouds heated by AGN ultraviolet continuum
and lying at a distance (e.g. Barvainis \cite{barvainis87})
$$ R = 0.8\ \left(L_{UV}\over 10^{11}\, \LSUN\right)^{1/2} 
      \left(T_{dust}\over 1000\,{\rm K} \right) ^{-2.8} 
\ \ \ {\rm pc} \eqno(1) $$
where $L_{UV}$ is the AGN luminosity in the UV and $T_{dust}$ is the
dust equilibrium temperature.

Interestingly,
the dust continua of Sy2's are systematically cooler than in Sy1's, but
the dust temperatures found in the type 2 Seyferts NGC5506 and NGC1068 are 
quite high ($\simeq$800 K) and only $\simeq$200 K cooler than those found 
in Sy1's. 
This implies that regions quite close ($\simeq$1 pc) to the nucleus 
are directly visible or, alternatively, that the
non--stellar continuum seen in these objects is scattered radiation.
This last possibility appears unlikely on energetic grounds:
the observed ratios $L_{nuc}(2.29)/L_{FIR}$ are within the range 
spanned by Sy1's (see last column of Table~\ref{tabcont}). More specifically,
if we assume that the intrinsic hot--dust spectrum of NGC1068 is the same (in 
luminosity and colour) to that of A0556-38, the efficiency of
the scattering ``mirror'' must be 6\%.  Adopting other Sy1 templates yields
more prohibitive values, e.g. 10\% and 20\% using IC4229a and NGC3783,
respectively.

The distance over which the dust equilibrium temperature drops from
1000 to 800 K follows from Eq.~(1) and is $\simeq$1 pc for 
$L_{UV}$=$10^{11}$ \LSUN.
Simple obscuration could account for the observed dust temperatures of 
NGC1068  
only by fine--tuning of the thickness and inclination of the torus.
%
%

\subsection{ CO(2,0) equivalent widths and starburst activity }
\label{COindex}

The strongly saturated CO(2,0) band--head  is primarily sensitive to
microturbulent photospheric motions ($\xi$) which increase at lower stellar
temperatures and are larger in supergiants than giants of a given temperature
(McWilliam \& Lambert \cite{lambert84}, \cite{OMO93} and references therein).
This indirect dependence on surface gravity is sometimes used to argue
for/against the presence of red supergiants in galaxies.
However, this index alone cannot be generally
used as a starburst tracer because a given equivalent width could be
equally well obtained with young stars of relatively low metallicity,
or with an old and highly metallic population dominated by cool red giants.
Adding measurements of other stellar features more sensitive to the stellar
temperature,
rather than $\xi$, may in principle help distinguishing between stellar
luminosity classes. However, the results obtained in \cite{OMO93} and 
\cite{paper1} indicate that equivalent widths alone
do not generally provide a convenient tool to trace supergiants.

In practice, the only objects which can be directly classified as young
stellar populations are those with \EW(2.29) significantly larger than
the maximum equivalent width measured in the most metallic 
(i.e. coolest) old systems known. 
Based on available data, a safe borderline could
be drawn at \EW(2.29)=15 \AA\ 
and 1 \AA\  above the values found in bright elliptical galaxies 
and in Terzan 1, a highly metallic globular
cluster in the bulge of our Galaxy (Origlia et al. \cite{origlia97}).\\
However, such a selection criterion is not necessarily useful for 
tracing red--supergiants.
On the one hand, many (but not all) of the young LMC clusters display
CO(2,0) indices lying well above the 15 \AA\  borderline
(see Table 1 of Oliva \& Origlia \cite{origlia98}).
On the other hand, well known starbursters often fall 
close or below this limit.
In particular, only 1/3 of the
starburst galaxies studied in \cite{paper1} 
display unmistakably strong CO(2,0) features
(see Table 2 of \cite{paper1}).  
Even more confusing is the case of NGC6240 which displays an
extremely strong CO(2,0) feature ($\simeq$18 \AA, see Table~\ref{tabew}) but
whose starburst origin has been sometimes questioned in the literature
(e.g. Lester \& Gaffney \cite{lester94}).

\subsection{Mass to light ratio and starburst activity}
\label{moverl}

As already discussed in Sect. 4.2 of \cite{paper1}, the observed velocity
dispersion ($\sigma$) can be directly related to the dynamical
mass of the system by assuming that the stars responsible for the IR emission
are moving in a dynamically relaxed isothermal sphere whose effective 
diameter is larger than the projected size of the spectrometer aperture.
Within these limits, 
the dynamical mass and the mass to light ratio ratio are simply given by
$$ M = 5.6\times10^8 
    \left(\sigma\over 100\, {\rm km/s}\right)^2
    \left(D\over 10\, {\rm Mpc}\right)
    \left(\theta_{obs}\over 5\arcsec\right) \ \ \ \MSUN \eqno(2) $$
$$ \MOL = {2.1
      \over F_{11}}
                \left(\sigma\over 100\, {\rm km/s}\right)^2
                \left(D\over 10\, {\rm Mpc}\right)^{-1}
                \left(\theta_{obs}\over 5\arcsec\right) \ \ \ \eqno(3) $$
where $\theta_{obs}$ is the spectrometer aperture, $D$ is the distance to the
galaxy, $\sigma$ the stellar velocity dispersion and
$F_{11}$ the observed stellar flux ($\times 10^{-11}$ erg
cm$^{-2}$ s$^{-1}$ $\mu$m$^{-1}$) corrected 
for the non--stellar (nuclear) contribution, as described in 
Sect.~\ref{nonstellar}, and de--reddened adopting an intrinsic stellar
colour H--K=0.2.
The value of \MOL\  is expressed in the conventional solar units, i.e.
normalizing the stellar $L_\lambda$  to 
$L_\odot$(H)=$4.48\times10^{32}$ erg s$^{-1}$ \MIC$^{-1}$,
the solar luminosity in the H band.

\begin{figure}[!]
\centerline{\resizebox{8.8cm}{!}{\rotatebox{0}{\includegraphics{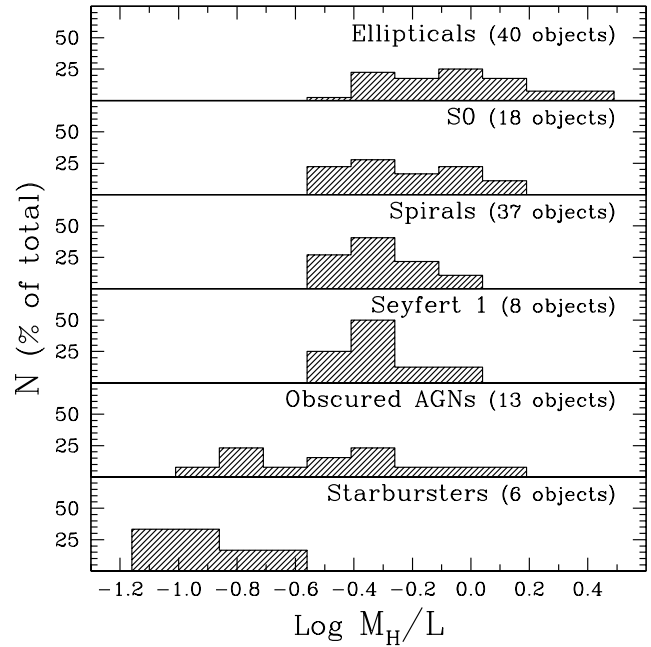}}}}
\caption{
Distribution of mass to light ratios inferred from the stellar flux and
velocity dispersions derived from the IR spectra.
Note that all the starbursters are clearly separated from normal
galaxies, while the distribution of Sy1's is very similar to that of
normal spirals (both have a mean value of 0.5). 
}
\label{figmol}
\end{figure}

\begin{table}[!t]
\caption{ Derived mass to light ratios}
\label{tabmol}
{
\def\UNO{\mbox{$^{(1)}$}}
\def\DUE{\mbox{$^{(2)}$}}
\def\TRE{\mbox{$^{(3)}$}}
\def\QUA{\mbox{$^{(4)}$}}
\def\CIN{\mbox{$^{(5)}$}}
\def\A{\rlap{$^a$}}
\def\B{\rlap{$^b$}}
\def\C{\rlap{$^c$}}
\def\D{\rlap{$^d$}}
\def\E{\rlap{$^e$}}
\def\SIM{\llap{$\sim$}}
\def\UL{\llap{$<$}}
\def\tablerule{\noalign{\hrule}}
\def\JM{\hglue-0.5em}
\def\J{\hglue0.5em}
\def\AVERA#1#2#3{&& && && && && &\cr\noalign{\vglue-18pt}
           && && \hrulefill && && \hrulefill && \hrulefill &\cr
           && && && && && &\cr\noalign{\vglue-12pt}
           && && #1 && && #2 && #3 &\cr}
\def\Skip{&& && && && && &\cr\noalign{\vglue-8pt}}
\def\SKIP{&& && && && && &\cr\noalign{\vglue-6pt}}
\def\BOT{\noalign{\vglue-6pt}&& && && && && &\cr\tablerule}
\hbox{\vbox{\tabskip=0pt\offinterlineskip
\halign to\hsize{
\strut#& \vrule #\tabskip=.2em plus3em&
#\hfil&#\vrule&       
\hfil#\hfil&#&        
\hfil#\hfil&#&        
\hfil#\hfil&#\vrule&  
\hfil#\hfil&\vrule    
# \tabskip=0pt\cr
\tablerule\SKIP
&& Object  && \hglue1em D\UNO \hglue0.5em && $F_*$(H)\DUE 
       && \hglue0.8em$\sigma$\TRE\hglue0.8em 
       && \hglue0.8em$M/L_{\rm H}$\QUA\hglue0.6em &\cr
\SKIP\tablerule\SKIP
&& {\it Type 1 Seyferts} && && && && &\cr
&& NGC~1365 && 20 && 7.3 && 151 && 0.33 &\cr
&& NGC~3783 && 37 && 3.1 && 152 && 0.43 &\cr
&& NGC~4593 && 37 && 5.4 && 209 && 0.47 &\cr
&& IC~4329a && 64 && 4.3 && 225 && 0.39 &\cr
&& Mkn~1095 && \JM130 && 1.1 && 177 && 0.45 &\cr
&& Mkn~1239  && 78 && 1.5 && \SIM250\B\ && \SIM1.0 &\cr
&& Fairall~9 && \JM186 && 1.0 && 228 && 0.60 &\cr
&& MCG-6-30-15 && 30 && 1.7 && 159 && 1.08 &\cr
&& A~0556-38 && \JM138 && 1.3 && 183 && 0.38 &\cr
\SKIP
&& {\it {\it Obscured AGNs$^{\;a}$} } && && && && &\cr
&& NGC~1052 &&  10 && 10. && 212 && 0.91 &\cr
&& {\bf NGC~1068} && 19 && 16. && 160 && {\bf 0.18} &\cr
&& NGC~2110 && 29   && 12. && 268 && 0.44 &\cr
&& NGC~2992 && 31 && 7.9 && 206 && 0.37 &\cr
&& NGC~4594 && 15 && 30. && 280 && 0.37 &\cr
&& NGC~4785 && 50 && 3.5 && 187 && 0.43 &\cr
&& {\bf NGC~4945} && \J4   && 52. && 134 && {\bf 0.18} &\cr
&& NGC~5506 && 24 && 4.2 && \SIM180\B\ && \SIM0.7  &\cr
&& {\bf NGC~6240} &&  98 && 10. && 300 && {\bf 0.18} &\cr
&& {\bf NGC~7582} && 19   && 20. && 157 && {\bf 0.13} &\cr
&& {\bf Circinus} &&  \J4   && 71. && 168 && {\bf 0.21} &\cr
&& A~0945-30 && 31  && 6.6 && 206 && 0.44 &\cr
&& M~87     &&  19   && 10. && 321 && 1.10 &\cr
\SKIP
&& {\it Starbursters} && && && && &\cr
&& {\bf NGC~253 } &&  \J4   && 56. && 109 && {\bf 0.11} &\cr
&& {\bf NGC~1614} && 62   && 9.2 && 150 && {\bf 0.08} &\cr
&& {\bf NGC~1808} && 10   && 25. && 154 && {\bf 0.20} &\cr
&& {\bf NGC~3256} && 35   && 9.3 && 127 && {\bf 0.10} &\cr
&& {\bf NGC~7552} &&  22   && 15. && 104 && {\bf 0.07} &\cr
&& {\bf NGC~7714} &&  40   && 3.7 && 117 && {\bf 0.19} &\cr
\BOT }}}

\smallskip
\def\NOTA#1#2
{\hbox{\vtop{\hbox{\hsize=0.03\hsize\vtop{#1}}}
      \vtop{\hbox{\hsize=0.95\hsize\vtop{#2}}}}}
 
\NOTA{\UNO}{ Adopted distance (Mpc).}
 
\NOTA{\DUE}{ Stellar flux at 1.65 $\mu$m 
[=$F_*$(1.62$\,\mu$m)/1.06, units of
10$^{-11}$ erg cm$^{-2}$ s$^{-1}$ $\mu$m$^{-1}$]
 corrected for reddening using the extinction
${\rm A_H}$ listed in Table 2.}
 
\NOTA{\TRE}{ Average dispersion velocity of stellar absorption features 
(km/s, see Table 1). }
 
\NOTA{\QUA}{ Mass to light ratio (solar units) derived from 
above quantities, see  Eq.~(1) and Sect.~\ref{moverl}. 
Galaxies with \MOL$<$0.30 are in bold type.}
 
\vskip2pt
\NOTA{$\;^a$}{  Objects with an obscured active nucleus,
classified from optical and/or X--ray spectra (see also 
Sect.~\ref{individual})}
 
\NOTA{$\;^b$}{ Stellar features are very diluted 
and $\sigma$ is uncertain. Mkn1239 is not included in Fig.~\ref{figmol}.}

}
\end{table}

The available data and derived mass to light ratios are summarized in 
Table~\ref{tabmol} while the distribution of \MOL\    in different types
of galaxies is displayed in Fig.~\ref{figmol} where we also include 
values for spirals and ellipticals from
IR photometric data and optical $\sigma$'s 
(Whitmore et al. \cite{whitmore}, Impey et al. \cite{impey}).

All normal galaxies have mass to light ratios in the range 0.3--3 and
consistent with a population of old (giant) stars plus a quiescent 
star formation activity (if any). 
The distribution of starbursters occupies only the region with \MOL$<$0.25 and
is clearly separated from that of normal galaxies. This demonstrate that
the value of \MOL\   provides an unambiguous tool to trace 
powerful starburst events.

The most striking result is that the distributions of Sy1's and normal
spirals are virtually indistinguishable (both have a mean value of 0.51), 
while about half 
of the obscured AGNs 
have much lower mass to light ratios and similar to those found in
genuine starbursters. This indicates that obscured AGNs have
experienced starburst activity within a region of a few arcsec
from the nucleus. 
Such an event  is likely to be old ($\ga$100 Myr) and extremely
powerful. Old, because its associated HII
regions (if any) are often not detected in hydrogen recombination lines 
(see \cite{paper1}).  Powerful, because the mass to light ratio
is significantly decreased only if the old starburst has produced 
enough red supergiant/AGB stars to out-shine the background emission
from the old population of RGB stars or, in practice, if 
the mass of gas  processed into stars
is at least a few \% of the galaxy mass. This also implies that, sometimes
in the past, the starburst was 1--2 orders of magnitude more
luminous and probably out--shone the AGN (if any), i.e. the galaxy
could be classified as a genuine starburster.
It is also interesting to note that Sy1 galaxies with known circumnuclear
HII regions (e.g. NGC1365) do not display unusually low \MOL\  ratios,
which simply indicates that the young starburst traced by the H
recombination lines is relatively weak. 
Similarly, the young/weak starbursts traced by the rings of HII regions
often found in Sy2's are not those responsible for the \MOL\  ratios
found here. Spectacular examples are NGC1068 and Circinus
which both exhibit HII rings with radii much larger than the regions
sampled in our spectra (see Neff et al. \cite{neff},  Marconi et al. 
\cite{marconi}).\\

\subsubsection{Possible caveats}

The mass to light ratios listed in Table~\ref{tabmol} and plotted
in Fig.~\ref{figmol} rely on the assumption that a simple
relationship between stellar velocity dispersion and dynamical mass,
i.e. Eq.~(2), could be used in all types of objects. 
Although many effects could in principle influence
the $\sigma$--mass relationship (e.g. the geometry and inclination of the 
system),
these should average out if a large enough sample of objects
is considered. In particular, no systematic effect should affect
the measurements of different types of Seyferts if these are really the 
same class of objects seen under different angles and if the orientation
of the obscuring torus is not related to the inclination of the galaxy,
i.e. if the AGN unification scheme is strictly correct.
The only observational bias which could introduce a systematic shift
of the Sy1 histogram in Fig.~\ref{figmol} is related to the fact that
half of the Sy1's in our sample are significantly more distant
than the unobscured AGNs.  
In these objects the dynamical mass -- and hence \MOL\  --
could be overestimated 
if the projected size of the spectrometer is larger than their effective radii.
However, available IR images (e.g. Kotilainen et al. \cite{koti1})
indicate that the stellar continuum of these galaxies is extended over 
many arcsec and should therefore fill the aperture of the spectrometer.
A definitive confirmation could come from spectra taken with narrower slits
which should yield smaller stellar fluxes ($F\!\propto\!\theta$) and 
similar mass to light ratios as those derived here.

\subsection{ Comments on individual sources }
\label{individual}

\subsubsection{ NGC4945}

This nearby spiral is a spectacular example of the AGN--starburst dichotomy.
At wavelengths shorter than 1 keV the object appears as a genuine starburster
and all the observational properties, including its FIR luminosity,
can be accounted for by starburst activity alone. Prior to the X--ray
observations discussed below, this galaxy
was usually classified as a starburster, a scheme which was also adopted
in \cite{paper1}.
When looked at in the X--rays, however, the galaxy becomes the
perfect example of an absorbed Seyfert 2 nucleus, i.e. 
powerful emission at $\ga$100 keV (NGC4945 is the brightest Sy2 in the
hard--X sky, Done et al. \cite{done96}) which rapidly fades at lower
energies where the AGN is also revealed by a prominent Fe--K line with
an equivalent width of $\simeq$1 keV.
The hard X--ray luminosity and $L$(100 keV)/$L(FIR)$ ratio are similar
to those found in the Circinus galaxy, a nearby Sy2 whose IRAS luminosity
is fully accounted for by the AGN alone
(Oliva et al. \cite{oliva99}).
It is therefore likely that the FIR luminosity of NGC4945  is also dominated
by the active nucleus.

\subsubsection{ NGC5506}
\label{ngc5506}

This Seyfert 2 galaxy is the only AGN in our sample with detectable Br13, Br14
hydrogen emission lines (see Fig.~\ref{figsy2}). The optical spectrum
is also characterized by prominent narrow lines over a relatively faint 
continuum. The equivalent width of \HA\  is $\simeq$400 \AA,  i.e.
among the largest observed in Sy2's 
(e.g. Morris \& Ward \cite{morris88}) and comparable to that found in
young starburst galaxies whose spectra are dominated by HII regions 
photoionized by massive O stars. 
However, the large strength of high excitation
lines ([NeV], [FeVII], HeII) argue for a non--stellar ionizing continuum.

The IR continuum is dominated by emission from hot dust
which strongly dilutes the absorption features even at 1.6
\MIC\  (see Fig.~\ref{figsy2}).
The inferred mass to light ratio, although uncertain, is compatible
with a normal (old) stellar population.

\subsubsection{ NGC6240}
\label{ngc6240}
This peculiar object is an interacting system with a bolometric
luminosity close to the limit of IRAS ultraluminous galaxies.
A deeply hidden AGN is revealed by X--ray spectra
showing a prominent FeK 6.4 KeV line
and powerful continuum emission at 100 keV (Iwasawa \& Comastri 
\cite{iwasawa98}, Iwasawa \cite{iwasawa99}, Matt \cite{matt99}).
The active nucleus is much less evident at longer wavelengths
but its contribution to the bolometric luminosity
is likely to be important
because the $L$(100 keV)/$L(FIR)$ ratio is similar to that measured in the
Circinus galaxy.

Several IR spectroscopic studies exist in the literature (Doyon et al.
\cite{doyon94}, Lester \& Gaffney \cite{lester94},
Shier et al. \cite{shier96}) 
reaching contradictory conclusions on the nature
of the red stellar population, i.e. young supergiants/AGB stars vs. old, 
low mass giants. 
In particular, the very large velocity
dispersion and relatively high mass to light ratio 
have been used to argue for 
an old elliptical--like system (e.g. Doyon et al. \cite{doyon94}).
However, these conclusions were drawn without correcting
the stellar flux for extinction which is indeed very high, about 1
magnitude at 1.65 \MIC\  (see Table~\ref{tabcont}).

The mass to light ratio derived here is unequivocally in the starburst 
domain and demonstrates that the near IR emission from
this massive object is still
dominated by a young stellar population. It should be noted that this
conclusion is still valid if the larger velocity dispersion quoted
in the above papers, i.e. $\sigma$=350 km/s, is adopted.

Another peculiarity of this object is the combination of very strong
CO(2,0), the deepest in our sample, and relatively 
shallow CO(6,3). The ratio \EW(2.29)/\EW(1.62)=4.1 is 30\% larger
than in starburst galaxies and young LMC clusters (see Table 1 and
Fig. 5 of Oliva \& Origlia \cite{origlia98}). 
The only other object
with a comparably large ratio is NGC330, a well known young and metal poor
cluster in the SMC. 
Since \EW(1.62) is much more sensitive to metallicity than \EW(2.29),
this indicates that NGC6240 is a relatively metal poor system as well.
Detailed modelling confirms this idea and yields a metallicity
$\simeq$1/10 solar, the lowest among the galaxies in our sample 
(see Oliva \& Origlia \cite{origlia98}).

\section{Conclusions}
\label{conclusion}

Spectra of active galaxies including the
stellar absorption features of Si at 1.59 \MIC, CO(6,3) at 
1.62 \MIC, and CO(2,0)
at 2.29 \MIC\  have been used to trace red supergiants and  thus
investigate the occurrence of old/exhausted circum--nuclear starbursts.
We find that the commonly used CO first overtone
spectroscopic/photometric index does not provide a reliable diagnostic
for distinguishing red supergiants, i.e. starbursts, from metallic red
giants, i.e. old stellar systems. 

A much more sensitive starburst tracer is the
mass to light ratio at 1.65 \MIC\ (\MOL) which is derived from the
observed stellar flux (de-reddened and corrected for non-stellar emission) 
and velocity dispersion of the stellar absorption features.
All starbursters old enough to have produced red supergiants do 
display values of \MOL\   at least 4 times smaller than normal 
elliptical/spirals.

In active galaxy nuclei, old and powerful starbursts are relatively common
in obscured AGNs  (5 objects out of 13) while absent in Seyfert 1's 
(0 objects out of 8). If confirmed on a larger sample, this result
may add support to AGN evolutionary models such as those
originally proposed to explain
the nature of ultraluminous IRAS galaxies and their possible relationship
with quasars (e.g. Sanders et al. \cite{sanders88}, Norman \& Scoville
\cite{norman88}, Heckman et al. \cite{heckman89}).
In this scenario circum--nuclear starburst
and nuclear activity are both triggered by gas accreting toward the nucleus
 which also accounts for the obscuration of the AGN at early
stages of its evolution.

Another interesting result is that the non--stellar continuum is very red
and compatible with emission from hot ($\simeq$1000 K) dust even in bare
Sy1's such as NGC3783 and IC4329a. Therefore, we do not confirm 
previous claims of much flatter nuclear spectra and find strong indication
that the featureless continuum in the near IR is due to reprocessed
AGN radiation in all types of Seyferts.
Moreover, the nuclear non--stellar continuum of Sy2's (when detected) is only
slightly cooler ($\simeq$800 K) than in Sy1's. 
The observed flux is too high to be accounted for by
scattered light and therefore indicates that the material obscuring the AGN
must have a quite small ($\la$1 pc)
projected size.

\begin{acknowledgements}
This work was partly supported by the Italian Ministry for University
and Research (MURST) under grant Cofin98-02-32.
We are grateful to the anonymous referee for his/her comments and critics 
which helped us to improve the quality of the paper.
\end{acknowledgements}

\end{document}